\documentclass[%
 reprint,
 amsmath,amssymb,
 aps,
pra,
]{revtex4-1}
\usepackage{xcolor}
\usepackage{physics}
\usepackage{graphicx}
\usepackage{dcolumn}
\usepackage{relsize}
\usepackage{comment}

\usepackage{bm}

\begin{document}

\preprint{APS/123-QED}

\title{
Entanglement degradation under local dissipative Landau-Zener noise}
\author{Melika Babakan}
\author{Arman Kashef}
\author{Laleh Memarzadeh}
\email{memarzadeh@sharif.edu}
\affiliation{%
Department of Physics, Sharif University of Technology, Tehran, Iran
}%
\date{\today}

\begin{abstract}
We study entanglement degradation when noise on one share of an entangled pair is described  by the dissipative  Landau-Zener model. We show that spin-coupling direction to the environment significantly affects entanglement dynamics. In particular, for zero bath temperature in the slow-driving regime with transversal coupling, entanglement remains intact and in the fast-driving regime transversal noise have less destructive affects on entanglement compared to the longitudinal noise. Furthermore, we show that non-adiabatic dynamic is more in favour of preserving entanglement compared to adiabatic evolution. 

\end{abstract}

\maketitle

\section{Introduction}
The essential role of entanglement in quantum information science and technology
\cite{Bennett1993, Ekert1991, Bouwmeester1997, Jozsa2003, yin_entanglement_based_2020} urges analysing its generation \cite{braun_creation_2002,
benatti_environment_2003,
krauter_entanglement_2011,memarzadeh_entanglement_2013}, distribution \cite{Hagley1997, Neumann2022} and protection against noise \cite{Maniscalco2008, Oliveir2008, Tong2010}. 
 In particular, analysing entanglement degradation when one share of an entangled pair experiences noise is essential due to its application in development of quantum networks 
\cite{kimble_quantum_2008, wehner_quantum_2018, Pompili2021}, 
quantum sensing
\cite{Degen2017}, quantum illumination \cite{Lloyd2008} and quantum communication protocols \cite{Briegel1998}. 

Regarding its importance, entanglement degradation has been studied from different aspects \cite{BenediktOmar2015, Paris2004, bellomo_entanglement_2010, rossi_degradation_2004, bullock_entanglement_2018, shaham_entanglement_2015, Gatto2019}. In \cite{shaham_entanglement_2015} the role of local noise on entanglement degradation is studied where noise is parametrized by local unital channels. 
A more realistic model is discussed in \cite{Gatto2019} where one share of an entangled pair is protected against noise and the other interacts with a thermal bath. The dynamics of this qubit is modeled by a Markovian master equation and role of coherent and incoherent parts of dynamics on entanglement degradation is discussed. 

\begin{figure}
    \centering
    \includegraphics[width=0.8\linewidth]{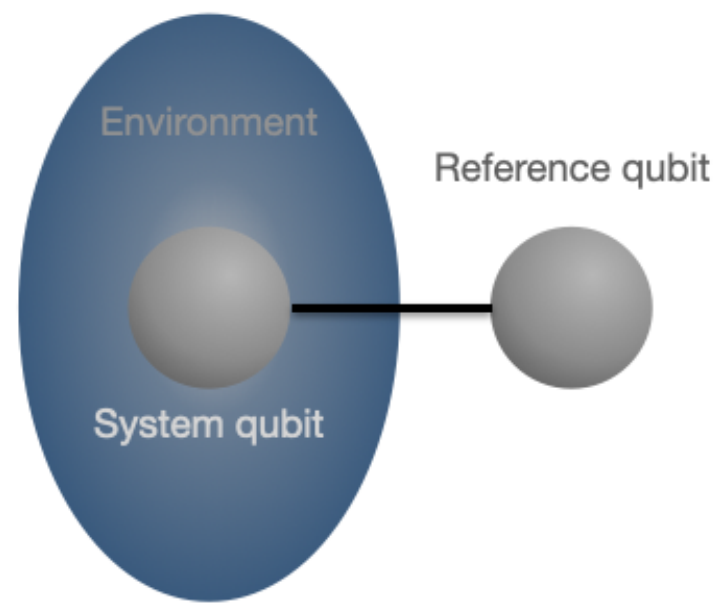}
    \caption{Schematic representation of two qubit system, where system qubit with Landau-Zener Hamiltonian interacts with environment in a thermal state and reference qubit does not experience noise.}
    \label{fig:system}
\end{figure}

Here we consider a setting as depicted in Fig.~(\ref{fig:system}),
where one share of an entangled pair $\ket{\Psi}_{SR}$, namely reference qubit ``R" is protected against noise and its local unitary evolution does not affect entanglement properties of the pair. System qubit ``S", interacts with a thermal bath and its coherent dynamics is generated by the time-dependent Landau-Zener Hamiltonian \cite{Landau1932, Zener1932}. Landau-Zener model is the simplest time-dependant model for addressing adiabatic/non-adiabatic transitions which has significant application in adiabatic quantum computation \cite{Farhi2000, Aharonov2007}. 
Also Landau-Zener serves as a successful model for describing the lowest energy levels employed in quantum annealing \cite{Santoro2006}. We consider the noise on the system qubit to be described by dissipative Landau-Zener model. In this model, we analyse the role of spin-coupling direction on entanglement degradation. 
Role of the spin-coupling direction in thermally assisted quantum annealing and the benefit of having transverse spin-coupling for ground state probability has been discussed in \cite{Arceci2017}. Here, for entanglement degradation under dissipative Landau-Zener model we show that larger noise coupling angle is beneficial for preserving entanglement in slow-driving regime with zero bath temperature and for a transverse coupling the dynamic is not entanglement breaking at all. In the fast-driving regime with zero bath temperature, with transversal noise entanglement decay is less than entanglement degradation with longitudinal noise.  We  also address the role of adiabaticity of the dynamics on entanglement degradation. We show that by going from fast to slow-deriving regime, entanglement break down is more serious. 
 
For our analysis we work in the weak coupling regime to have Markovian dynamics. Deriving the master equation for Landau-Zener Hamiltonian (like any other time-dependant Hamiltonian), requires taking care of further subtleties. We use instantaneous eigenstate of Landau-Zener Hamiltonian and the approach given in \cite{ogawa} to derive the time-dependent generator of the dissipative Landau-Zener model. As discussed in \cite{ogawa} when temporal change timescale of system Hamiltonian and relaxation timescale is much larger than intrinsic beat timescale, the derived master equation in this approach, is valid both in adiabatic and non-adiabatic regime. Hence, by proper choice of parameters we can discuss the role of adiabaticity of dynamics on entanglement degradation. The other challenging issue is solving the master equation with time-dependant generator. The analytical solution does not exists unless in some particular range of parameters. Hence, beside providing analytical  solution in the slow-driving regime, we use numerical techniques to explore a wider range of parameters and extend our analysis to fast-driving regime. 

The structure of the paper is as follows. In \S\ref{sec:model} after providing the description of a single qubit dissipative Landau-Zener model, we derive the master equation and discuss the proper range of the parameters for having a valid Markov master equation. We discuss the evolution of a general two-qubit system under local dissipative Landau-Zener noise in \S\ref{sec:TPME}. Then we proceed to analyze entanglement dynamics in \S\ref{sec:EntDyn} and report our results in slow and fast-driving regimes. We discuss the results and conclude in \S\ref{sec:conclusion} with an outlook for potential future work.

\section{Single qubit dissipatative Landau-Zener model}
\label{sec:model}
In this section, we describe the dynamics of a single qubit with Landau-Zener Hamiltonian that interacts with a thermal bath. In \S\ref{subsec:TDMD} we use instantaneous eigenstates of Hamiltonian and follow the approach of \cite{ogawa} to derive the master equation in the weak coupling limit. In order to have a compact representation for Lindblad operators, we need to work in a rotated basis which we explain in \S\ref{sec:TDMD-RB}. As entanglement is invariant under change of basis, this time-dependant rotated frame enables us to obtain analytical results in a slow-driving regime and also have numerical analysis of entanglement dynamics in fast-driving regime. 

The Landau-Zener Hamiltonian which describes the coherent dynamics of system is give by
\begin{equation}
\label{eq:HA}
    H_S(t)=\hbar(\Delta\sigma_x+vt\sigma_z),
\end{equation}
where $\sigma_x=\begin{pmatrix}0 & 1\cr 1 &0\end{pmatrix}$ and $\sigma_z=\begin{pmatrix}
	1&0\cr 0&-1
\end{pmatrix}$ are Pauli operators and $\Delta, v$ are positive real parameters of the Hamiltonian. From here on, we set Planck constant $\hbar=1$. Energy levels of system and the corresponding eigenvectors are given by 
\begin{equation}
\label{eq:HASpectrum}
   \epsilon_{\pm}(t)=\pm\Omega(t),\;\ket{\epsilon_{\pm}(t)}= \frac{1}{N_\pm(t)}
   \begin{pmatrix}  \Delta  \\  \pm \Omega(t)-vt \\ \end{pmatrix},
\end{equation}
with $N_\pm(t)=\sqrt{2\Omega(t)(\Omega(t)\mp vt)}$ and 
\begin{equation}
\label{eq:Omega}
    \Omega(t)=\sqrt{v^2t^2+\Delta^2}.
\end{equation}


To model noise on the system of study, we assume that system qubit interacts with a thermal environment described by the following Hamiltonian
\begin{equation}
    H_E=\int_0^{\omega_{\rm max}} d\nu  b^{\dagger}(\nu)b(\nu),
\end{equation}
where $b(\nu)$ and $b(\nu)^{\dagger}$ are bosonic operators of environment $[b(\nu),b^{\dagger}(\nu')]=\delta(\nu-\nu')$ and $\omega_{\rm max}$ is the cut-off frequency of the harmonic oscillators in the bath. 
The total Hamiltonian of system and environment is given by
\begin{equation}
    H_{\rm total}(t)=H_{S}(t)+H_E+\lambda H_{SE},
\end{equation}
where $\lambda>0$ is a constant denoting the strength of the interaction between the system and the environment and $H_{SE}$ describes the interaction between system and environment:
\begin{equation}
\label{eq:HAE}
    H_{SE}=\mathbf{A}\otimes\mathbf{B},
\end{equation}
where
\begin{align}
    &\mathbf{A}=\frac{1}{2}(\cos(\theta)\sigma_z+\sin(\theta)\sigma_x),\cr
    &\mathbf{B}=\int_0^{\omega_{\rm max}}d\nu g(\nu)(b(\nu)+b^{\dagger}(\nu)).
    \label{eq:AA}
\end{align}
Here $g(\nu)$ is the coupling function and $\theta\in[0,2\pi)$ determines the spin-coupling direction. By varying $\theta$ from $0$ to $\frac{\pi}{2}$, the spin-coupling changes from  longitudinal to transversal.

\subsection{Master equation in the weak-coupling limit}
\label{subsec:TDMD}
In this subsection, we derive the master equation describing the dynamics of system qubit with Landau-Zener Hamiltonian in Eq.~(\ref{eq:HA}) that interacts with a thermal bath with interaction Hamiltonian given in Eq.~(\ref{eq:HAE}). For rigorous construction of Markov quantum master equation we follow the formalism in \cite{ogawa} and discuss the range of parameters that guarantees the validity of the derived Markov master equation for desciribg the dynamics even in non-adiabatic regime \cite{ogawa}.

Following the approach described in \cite{ogawa} we derive the following quantum master equation: 
\begin{align}
\label{eq:GenMaster}
    \dot{\rho}_S(t)&=\mathcal{L}_t[\rho_S(t)],\cr
    \mathcal{L}_t[\bullet]&=-i[H_S(t)+\lambda^2 H_{LS}(t),\bullet]+\lambda^2\mathcal{D}_t[\bullet],\cr
   \mathcal{D}_t[\bullet]&=\sum_{m}\gamma(\omega_m(t))(A(\omega_m(t))\bullet A^\dagger(\omega_m(t))\cr
    &-\frac{1}{2}\{A^\dagger(\omega_m(t))A(\omega_m(t)),\bullet\}), \cr
\end{align}
where for deriving Lindblad operators $A(\omega_m(t))$ instantaneous eigensates of Hamiltonian $H_S(t)$ are used:
\begin{align}
\label{eq:AOriginal}
    &A(\omega_1(t))=\Pi_{\epsilon_-}(t)\textbf{A}\Pi_{\epsilon_+}(t),\cr
    &A(\omega_2(t))=\Pi_{\epsilon_+}(t)\textbf{A}\Pi_{\epsilon_-}(t),\cr
    &A(\omega_3(t))=\Pi_{\epsilon_-}(t)\textbf{A}\Pi_{\epsilon_-}(t)+\Pi_{\epsilon_+}(t)\textbf{A}\Pi_{\epsilon_+}(t).
\end{align}
Here $\Pi_{\epsilon_{\pm}}=\ket{\epsilon_\pm(t)}\bra{\epsilon_\pm(t)}$ are respectively projectors to eigenstates $\ket{\epsilon_{\pm}(t)}$ in Eq.~(\ref{eq:HASpectrum}) and $\omega_m(t)$ denotes transitions frequencies:
\begin{equation}
	\label{eq: TransitionFreq}
    \omega_1(t)=-\omega_2(t)=2\Omega(t),\;\;
    \omega_3(t)=0.
\end{equation}
Following the description in appendix  \S\ref{sec:coef} we have:
\begin{align}
\label{eq:Kossakowski}
\gamma(\omega_m(t))&=2\pi J(\abs{\omega_m(t)})\Theta(\omega_m(t))(\bar{n}(\abs{\omega_m(t))})+1)\cr
&+2\pi J(\abs{\omega_m(t)})\Theta(-\omega_m(t)))\bar{n}(\abs{\omega_m(t)}))\cr
&+
2\pi T \delta(\omega_m(t))
\end{align}
where $m=1, 2, 3$ addresses different transition frequencies in Eq.~(\ref{eq: TransitionFreq}), $\Theta(x)$ is the step function and
\begin{equation}
    \bar{n}(\omega)=\frac{1}{e^{\frac{\omega}{T}}-1},
    \label{eq:meann}
\end{equation}
is the environment's mean photon number with frequency $\omega$ when its temperature is $T$ (we set Boltzmann constant $\beta=1$).
In Eq.~(\ref{eq:Kossakowski}) 
$J(\omega_m(t))=g^2(\omega_m(t))$ is the spectral density of the environment which is assumed to be be Ohmic:
\begin{equation}
    J(\omega_m(t))=\omega_m(t)e^{-\frac{\omega_m(t)}{\omega_c}},
    \label{eq:j}
\end{equation}
where $\omega_c$ is the cutoff frequency. 
Lamb-shift Hamiltonian in Eq.~(\ref{eq:GenMaster}) is the correction to Hamiltonian arising from weak coupling approximation, and is given in terms of Lindblad operators:
\begin{equation}
    \label{eq:LS}
    H_{LS}(t)=\sum_{m=1}^3 \mathcal{S}(\omega_m(t))A^{\dagger}(\omega_m(t))A(\omega_m(t)),
\end{equation}
For $\mathcal{S}(\omega_m(t))$ see appendix \S\ref{sec:coef}.
By using the formal definition of Lindblad operators in Eq.~(\ref{eq:AOriginal}), the explicit expression of Lindblad operators are given by matrices with time-dependent elements. Such a representation is not suitable for the aim of analyzing entanglement degradation. 
In the next subsection, we explain how to derive a compact explicit form of Lindblad operators for a qubit with Landau-Zener Hamiltonian interacting with a thermal bath according to the interaction Hamiltonian $H_{SE}$ in Eq.~(\ref{eq:HAE}).

It is worth mentioning that the Markovian master equation in Eq.~(\ref{eq:GenMaster}) is obtained for the range of parameters when secular approximation is valid. In order to guarantee the validity of secular approximation temporal change timescale of $H_A(t)$, intrinsic beat timescale and relaxation timescale, which are respectively denoted by $\tau_A(t)$, $\tau_S(t)$ and $\tau_R(t)$ must satisfy the following inequalities \cite{ogawa}:
\begin{align}
    &\tau_S(t)\ll\tau_R(t),\cr
    &\tau_S(t)\ll\tau_A(t).
    \label{eq:Secularapp}
\end{align}
These timescales for our model are given by
\begin{align}
&\tau_S(t)=\frac{1}{2\Omega(t)},\\
&\tau_R(t)=\frac{1}{2\pi J(\omega_1(t))(\bar{n}(\omega_1(t))+1)},
\end{align}
and
\begin{equation}
\tau_A(t) = 
     \begin{cases}
      \frac{2\Omega^2(t)}{v\Delta}&t\leq\frac{\Delta}{v}\\
      \\
       \frac{\Omega^2(t)}{v^2t} & t>\frac{\Delta}{v}
     \end{cases}
\end{equation}
In proceeding sections, when we provide relevant figures for discussing the role of different parameters in entanglement degradation, 
we choose parameters such that the inequalities in Eq.~(\ref{eq:Secularapp}) hold.

\subsection{Time-dependant master equation in rotated basis}
\label{sec:TDMD-RB}
In this subsection, we derive the explicit compact form for the generator of dynamic $\mathcal{L}_t$ in Eq.~(\ref{eq:GenMaster}) in a rotated basis \cite{Nalbach2014}. This generator will be used in the following sections for describing local Landau-Zener dissipative dynamics. 

In order to derive a compact form for Lindblad operators we use instantaneous eigenstates of  $H_S(t)$ in Eq.~(\ref{eq:HA}). Unitary time-dependent transformation that diagonalizes $H_S(t)$ is given by
\begin{equation}
\label{eq:rotate}
\mathcal{R}(t)=e^{i\phi(t)\sigma_y},
\end{equation}
where $\sigma_y=\begin{pmatrix}
	0&-i\\
	i&0
\end{pmatrix}$ is the second Pauli matrix and 
\begin{equation}
    \phi(t)=\frac{1}{2}\arctan(\frac{\Delta}{vt}). 
    \label{eq:phit}
\end{equation}
We denote any operator $O$ in the rotated basis by $\tilde{O}(t)$:
\begin{equation}
    \tilde{O}(t)=\mathcal{R}(t)O\mathcal{R}^{\dagger}(t).
\end{equation}
Therefore system Hamiltonian in the rotated basis is given by
\begin{equation}
    \tilde{H}_S(t)=\mathcal{R}(t)H_S(t)\mathcal{R}^\dagger(t)=\Omega(t)\sigma_z,
\label{eq:tildeHLZ}    
\end{equation}
where $\Omega(t)$ is defined in Eq.~(\ref{eq:Omega}). From Eq.~(\ref{eq:tildeHLZ}) it is clear that $\tilde{H}_S(t)$ is diagonal for all time $t$. 
The master equation in Eq.~(\ref{eq:GenMaster}) in the rotated basis is given by
\begin{align}
     \dot{\tilde{\rho}}(t)&=\tilde{\mathcal{L}}_t[\tilde{\rho}(t)],\cr
     \tilde{\mathcal{L}}_t[\bullet]&=-i[\tilde{H}_S(t)+\lambda^2\tilde{H}_{LS}(t)+\dot\phi(t)\sigma_y,\bullet]   +\lambda^2\tilde{\mathcal{D}}_t[\bullet],\cr
     \tilde{\mathcal{D}}_t[\bullet]&=\sum_m \gamma(\omega_m(t))(\tilde{A}(\omega_m(t))\bullet \tilde{A}^{\dagger}(\omega_m(t))\cr
    &-\frac{1}{2}\{\tilde{A}^{\dagger}(\omega_m(t)) \tilde{A}(\omega_m(t)),\bullet\}), 
     \label{eq:rotatemaster}
 \end{align}
where Lindblad operators in the rotated basis are given by
\begin{align}
     &\tilde{A}(\omega_1(t))=\tilde{\Pi}_{\epsilon_-}\tilde{\textbf{A}}(t)\tilde{\Pi}_{\epsilon_+},\cr
    &\tilde{A}(\omega_2(t))=\tilde{\Pi}_{\epsilon_+}\tilde{\textbf{A}}(t)\tilde{\Pi}_{\epsilon_-},\cr
    &\tilde{A}(\omega_3(t))=\tilde{\Pi}_{\epsilon_-}\tilde{\textbf{A}}(t)\tilde{\Pi}_{\epsilon_-}+\tilde{\Pi}_{\epsilon_+}\tilde{\textbf{A}}(t)\tilde{\Pi}_{\epsilon_+}. 
\end{align}
Projector operators $\tilde{\Pi}_{\epsilon_{\pm}}$ in rotated basis have a simple form: 
$\tilde{\Pi}_{\epsilon_{-}}=\ket{0}\bra{0}$ and $\tilde{\Pi}_{\epsilon_{+}}=\ket{1}\bra{1}$. In this basis, the Lindblad operators are derived as a function of time multiplied by a constant time-independent operators $\sigma_{\pm}=\frac{\sigma_x\pm i\sigma_y}{2}$ and $\sigma_z$:
\begin{align}
\label{eq:tildeLindOp}
     &\tilde{A}(\omega_1(t))=\frac{1}{2}\sin(\theta-2\phi(t))\sigma_-,\\
\label{eq:tildeLind1p}     
     &\tilde{A}(\omega_2(t))=\frac{1}{2}\sin(\theta-2\phi(t))\sigma_+,\\
\label{eq:tildeLind2p}
     &\tilde{A}(\omega_3(t))=\frac{1}{2}\cos(\theta-2\phi(t))\sigma_z.
\end{align}
We find the Lamb shift Hamiltonian in Eq.~(\ref{eq:LS}) in the rotated basis by using the explicit form of Lindblad operators in Eqs.~(\ref{eq:tildeLindOp})-(\ref{eq:tildeLind2p})
\begin{align}
&\tilde{H}_{LS}(t)\cr
&=\frac{1}{2}\sin^2(\theta-2\phi(t))\left(\mathcal{S}(\omega_1(t))-\mathcal{S}(\omega_2(t))\right)(\operatorname{id}+\sigma_z)\cr
&+\frac{1}{4}\cos^2(\theta-2\phi(t))\operatorname{id}.
\label{eq:tildeHLS}
\end{align}
Using the explicit form of system Hamiltonian in Eq.~(\ref{eq:tildeHLZ}), Lindblad operators in Eqs.~(\ref{eq:tildeLindOp})-(\ref{eq:tildeLind2p}) and the Lamb-shift term in Eq.~(\ref{eq:tildeHLS}), yields the explicit form of dynamic's generator in the rotated basis 
\begin{align}
\label{eq:Lt}
        \tilde{\mathcal{L}}_t[\bullet]&=-i[k(t)\sigma_z-\dot\phi(t)\sigma_y,\bullet]\cr
        &+f(t)(\sigma_-\bullet \sigma_+ - \frac{1}{2}\{\sigma_+\sigma_-,\bullet\})\cr
        &+g(t)(\sigma_+\bullet \sigma_- - \frac{1}{2}\{\sigma_-\sigma_+,\bullet\})\cr
        &+l(t)(\sigma_z\bullet\sigma_z-\bullet),
\end{align}

where $k(t)$, $f(t)$, $g(t)$ and $l(t)$ are given by
\begin{align}
  \label{eq:coefkfg}
  &k(t)=\Omega(t)+\frac{\lambda^2}{2}\sin^2(\theta-2\phi(t))(\mathcal{S}(\omega_1(t))-\mathcal{S}(\omega_2(t))),\\
  \label{eq:coeff}
   &f(t)=\frac{\lambda^2}{4}\sin^2(\theta-2\phi(t))\gamma(\omega_1(t)),\\
  \label{eq:coefg}
   &g(t)=\frac{\lambda^2}{4}\sin^2(\theta-2\phi(t))\gamma(\omega_2(t)),\\
   \label{eq:coefl}
  &l(t)=\frac{\lambda^2}{4}\cos^2(\theta-2\phi(t))\gamma(\omega_3(t)).
\end{align}

In the next section, we use this explicit form of dynamic's generator in Eq.~(\ref{eq:Lt}) for deriving the local Markovian master equation in a bipartite system. 

\section{Local dissipative Landau-Zener model}
\label{sec:TPME}
In this section, we describe local dissipative Landau-Zener model. We derive the master equation governing the dynamics of bipartite system where system qubit experience noise and reference qubit remains invariant. We derive the set of coupled first order differential equations for coefficeints of a general bipartite input state. For a specific range of parameters, we solve this set of coupled first order differential equations analytically for a one parameter initial pure state for which by varying the entanglement interpolates between zero and one.  

We consider a pair of entangled qubits. As depicted in Fig.~(\ref{fig:system}) in the setting under study, the reference qubit is protected against noise, therefore the dissipative dynamics of system qubit is the only source of noise that affects the entanglement between reference and system qubits. We denote the system-reference bipartite density matrix at time $t$ with $\rho_{SR}(t)$. Following the discussion of \S\ref{sec:model}, when system undergoes a dissipative Landau-Zener dynamics with system-environment interaction as described in Eq.~(\ref{eq:HAE}) and the reference qubit does not evolve, the dynamics of the pair is generated by $\mathcal{L}_t\otimes\operatorname{id}$ with $\mathcal{L}_t$ defined in Eq.~(\ref{eq:GenMaster}). 
Following the arguments in \S~\ref{sec:TDMD-RB}, to derive the explicit form of Lindblad operators as in Eqs.~(\ref{eq:tildeLindOp}), (\ref{eq:tildeLind1p}) and (\ref{eq:tildeLind2p}), it is required to rotate the basis by time-dependant unitary operator $\mathcal{R}(t)$ given in Eq.~(\ref{eq:rotate}). 
As we aim to analyze entanglement dynamics of bipartite system described by density matrix $\rho_{SR}(t)$ and entanglement is invariant under local unitary operations, we work in the rotated basis and analyze entanglement dynamics of
\begin{equation}
\label{eq:LocalRotation}
    \tilde{\rho}_{SR}(t)=(\mathcal{R}(t)\otimes\operatorname{id})\rho_{SR}(t)(\mathcal{R}^\dagger(t)\otimes\operatorname{id}).
\end{equation}
Therefore, in what follows, we focus on solving the following master equation in the rotated basis:
\begin{equation}
    \dot{\tilde{\rho}}_{SR}(t)=(\tilde{\mathcal{L}}_t\otimes \operatorname{id})\tilde{\rho}_{SR}(t),
    \label{eq:masterab}
\end{equation}
where $\tilde{\mathcal{L}}_t$ is given in Eq.~(\ref{eq:Lt}). To solve Eq.~(\ref{eq:masterab}) for density matrix $\tilde{\rho}_{SR}(t)$, we take it into account that any bipartite density matrix of $\tilde{\rho}_{SR}(t)$ can be written as:
\begin{align}
    &\tilde{\rho}_{SR}(t)=\frac{1}{4}\times\cr
    &\left(\operatorname{id} \otimes\operatorname{id} +\vec{s}(t).\sigma\otimes \operatorname{id}+\operatorname{id}\otimes\vec{r}(t).\sigma+\sum_{i,j=1}^3\chi_{ij}(t)\sigma_i \otimes\sigma_j\right), \cr
    \label{eq:rhoab}
\end{align}
with 
\begin{align}
    &s_i(t)=\operatorname{Tr}((\sigma_i\otimes\operatorname{id})\tilde{\rho}_{SR}(t)),\cr
    & r_i(t)=\operatorname{Tr}((\operatorname{id}\otimes\sigma_i)\tilde{\rho}_{SR}(t)),\cr
    &\chi_{ij}=\operatorname{Tr}((\sigma_i\otimes\sigma_j)\tilde{\rho}_{SR}),\cr
\end{align}
where $\sigma_i$s are Pauli operators. 
Furthermore, by 
using $\tilde{\mathcal{L}}_t$ in Eq.~(\ref{eq:Lt}) we have:
\begin{align}
\label{eq:tildeLPauliPauli}
    \tilde{\mathcal{L}}_t[\operatorname{id}] &=-2a_-(t)\sigma_z,\cr
    \tilde{\mathcal{L}}_t[\sigma_x]
    &=2k(t)\sigma_y-b(t)\sigma_x+2\dot{\phi}(t)\sigma_z,\cr
    \tilde{\mathcal{L}}_t[\sigma_y]&=
    -2k(t)\sigma_x-b(t)\sigma_y,\cr
    \tilde{\mathcal{L}}_t[\sigma_z]&=
    -2\dot{\phi}(t)\sigma_x-2a_+(t)\sigma_z,\cr
\end{align}
with
\begin{align}
    \label{eq:coeffs}
    &a_{\pm}(t)=\frac{1}{2}(f(t)\pm g(t)),\cr
    &b(t)=a_+(t)+2l(t).
\end{align}
Hence, by comparing the left and right hand sides of Eq.~(\ref{eq:masterab}) for $\tilde{\rho}_{SR}(t)$ as in Eq.~(\ref{eq:rhoab}) and considering the action of $\tilde{\mathcal{L}}_t$ on identity and Pauli operators in Eq.~(\ref{eq:tildeLPauliPauli}) we find a set of coupled first order differential equations for $\vec{s}(t)$, $\vec{r}(t)$ and $\chi(t)$. For vector $\vec{r}(t)$ it results that
\begin{equation}
\label{eq:rt}
 \forall t>t_{\rm int}\;\;\;\;   \vec{r}(t)=\vec{r}(t_{\rm int}),
\end{equation}
where $t_{\rm int}$ denots the initial time of the dynamics.
Furthermore, for $\vec{s}(t)$ and $\chi_{ij}(t)$ the following differential equations hold:
\begin{align}
    &\frac{d}{dt}\vec{s}(t)=Q(t)\vec{s(t)}+\vec{q}(t),\cr
    &\frac{d}{dt}\vec{\chi}_j(t)=Q(t)\vec{\chi}_j(t),\;\; j=1,2,3,
    \label{eq:hard1}
\end{align}
where
\begin{equation}
    \vec{\chi}_j(t)=\begin{pmatrix}
    \chi_{1j}(t)\\
    \chi_{2j}(t)\\
    \chi_{3j}(t)\\
    \end{pmatrix},\;\;
    \vec{q}(t)=\begin{pmatrix}
    0\\
    0\\
    -2a_-(t)\\
    \end{pmatrix},
\end{equation}
and 
\begin{equation}
    Q(t)=\begin{pmatrix}
    -b(t)&-2k(t)&-2\dot{\phi}(t)\\
    2k(t)&-b(t)&0\\
    2\dot{\phi}(t)&0&-2a_+(t)\\
    \end{pmatrix}.
    \label{eq:Qqt}
\end{equation}
Differential equations in Eq.~(\ref{eq:hard1}) have time-dependent coefficients. Therefore solving Eq.~(\ref{eq:hard1}) analyticaly and expressing the solution in a compact form is not possible. Hence, we analytically solve Eq.~(\ref{eq:hard1}) in a particular range of parameters where $v\to 0$ and $\Delta$ is finite for $t\ll\frac{\Delta}{v}$. For the rest of the range of parameters we solve the master equation in Eq.~(\ref{eq:masterab}) numerically. 

For analytical solution, first we discuss the limit of elements of matrix $Q(t)$ in Eq.~(\ref{eq:Qqt}) in the limit of $v\to 0 $ and $t\ll \frac{\Delta}{v}$ with finite $\Delta$. From Eq.~(\ref{eq:Omega}) we conclude that 
for $t\ll \frac{\Delta}{v}$, $\Omega(t)$ approaches $\Delta$. On the other hand
from Eq.~(\ref{eq:phit}) it is easy to see that
\begin{equation}
    \dot{\phi}(t)=-\frac{\Delta v}{2\Omega^2(t)}.
    \label{eq:phi}
\end{equation}
Therefore for $v\to 0$ and $t\ll \frac{\Delta}{v}$, $\dot{\phi}(t)$ approaches zero. Also it is clear from Eq.~(\ref{eq:phit}) that in this regime  $\phi(t)\approx\frac{\pi}{4}$. Therefore, from Eqs.~(\ref{eq:coefkfg})-(\ref{eq:coefl}) we conclude that $k(t)$, $f(t)$, $g(t)$ and $l(t)$ in this regime are respectively given by
\begin{align}
\label{eq:kfgl}
    &k= \Delta+\frac{\lambda^2}{2}\cos^2(\theta)(\mathcal{S}(2\Delta)-\mathcal{S}(-2\Delta)),\cr
    &f=\frac{\lambda^2}{4}\cos^2(\theta)\gamma(2\Delta),\cr
    &g=\frac{\lambda^2}{4}\cos^2(\theta)\gamma(-2\Delta),\cr
    &l=\frac{\lambda^2}{4}\sin^2(\theta)\gamma(0).
\end{align}
Therefore in this regime for coefficients $a_{\pm}(t)$ and $b(t)$ in Eq.~(\ref{eq:coeffs}) we have:
\begin{align}\label{eq:apmb}
&a_{\pm}=\frac{\lambda^2}{4}\cos^2(\theta)(\gamma(2\Delta)\pm(\gamma(-2\Delta)),\cr
&b=a_++2l
\end{align}
Hence, in this regime $Q(t)$ in Eq.~(\ref{eq:Qqt}) is time-independent. Furthermore, as $\dot{\phi}(t)\approx 0$, matrix $Q(t)$ is block-diagonal in this regime. Therefore it is possible to solve coupled differential equations in Eq.~(\ref{eq:hard1}) analytically and the solution of differential equations in Eq.~(\ref{eq:hard1}) in the regime of $v\to 0 $ and finite $\Delta$ for $t\ll \frac{\Delta}{v}$ is summarized as follows:
\begin{equation}
\label{eq:st}
\vec{s}(t)=\frac{1}{\sqrt{2}}\begin{pmatrix}
    e^{-b(t-t_{\rm int})}(A^+_0e^{2ik(t-t_{\rm int})}+A^{-}_0e^{-2ik(t-t_{\rm int})})\\
    ie^{-b(t-t_{\rm int})}(A^+_0e^{2ik(t-t_{\rm int})}-A^{-}_0e^{-2ik(t-t_{\rm int})})\\
    \sqrt{2}(B_0e^{-2a_+(t-t_{\rm int})}+\frac{a_-}{a_+})\\
    \end{pmatrix},
\end{equation}
and
\begin{equation}
\label{eq:chit}
\vec{\chi}_j(t)=\frac{1}{\sqrt{2}}\begin{pmatrix}
    e^{-b(t-t_{\rm int})}(A^+_je^{2ik(t-t_{\rm int})}+A^-_je^{-2ik(t-t_{\rm int})})\\
    ie^{-b(t-t_{\rm int})}(A^+_je^{2ik(t-t_{\rm int})}-A^-_je^{-2ik(t-t_{\rm int})})\\
    \sqrt{2}B_je^{-2a_+(t-t_{\rm int})}\\
    \end{pmatrix},
\end{equation}
where $A^{\pm}_j$, $B_j$ with $j=0,1,2,3$ are constant coefficients which are determined by the initial state: 
\begin{align}
    &A^{\pm}_0=\frac{1}{\sqrt{2}}(s_1(t_{\rm int})\mp i s_2(t_{\rm int})),\;\;\;\;\; B_0=(s_3(t_{\rm int})-\frac{a_-}{a_+})\cr
    &A^{\pm}_j=\frac{1}{\sqrt{2}}(\chi_{1j}(t_{\rm int})\mp i\chi_{2j}(t_{\rm int})), \; B_j=\chi_{j3}( t_{\rm int})
\end{align}
The analytical solution for a regime of $v\to 0$ and $t\ll\frac{\Delta}{v}$ as given in Eqs.~(\ref{eq:st}) and (\ref{eq:chit}) are used in the next section for analytical analysis of entanglement degradation under local dissipative Landa-Zener evolution. 
\section{Entanglement decay}\label{sec:EntDyn}
In this section, we analyse entanglement degradation in a bipartite qubit system when one share of an entangled pair undergoes dissipative Landau-Zener evolution. After general review on the measure of the entanglement we use, we analyse slow and fast-deriving regimes separately. Starting with a maximally entangled state, in the slow-driving regime we provide an analytical expression for entanglement in time. We discuss the dependence of  entanglement survival time to bath temperature and also spin coupling direction to the environment when bath temperature is zero. In the fast-deriving regime when bath temperature is zero we discuss entanglement behaviour in time for different range of parameters. Furthermore, we address the role of adiabaticity in entanglement degradation. 

To analyse entanglement between system and reference qubits we use negativity as a measure of entanglement. Like all entanglement measures, negativity is invariant under local unitary operations. Therefore, by taking into account Eq.~(\ref{eq:LocalRotation}) we have
\begin{equation}
    \mathcal{N}(\tilde{\rho}_{SR}(t))=\mathcal{N}(\rho_{SR}(t)).
\end{equation}
Hence to address entanglement between system and reference qubits, we focus on entanglement in state $\tilde{\rho}_{SR}(t)$. 
By definition, negativity of the bipartite qubit density matrix $\tilde{\rho}_{SR}(t)$ is given by \cite{VidalWerner2002}:
\begin{equation}
\label{eq:negativity}
    \mathcal{N}(\tilde{\rho}_{SR}(t)):=\frac{1}{2}\sum_{i=1}^4 (|\mu_i(t)|-\mu_i(t)),
\end{equation}
where $\mu_i(t)$s are eigenvalues of $\tilde{\rho}^{\top_R}_{SR}(t)$ which is the partial transposed of $\tilde{\rho}_{SR}(t)$ with respect to reference qubit $R$. For maximally entangled states, like bell state, negativity reaches its maximum value $\frac{1}{2}$ and for separable states it is zero. 

We consider a general pure bipartite state in its schmidt decomposition \cite{Peres93} for initial state:
\begin{equation}
\ket{\phi}=\cos{\eta}\ket{00}+\sin{\eta}\ket{11}.
\label{eq:initialState}
\end{equation}
By varying $\eta\in[0,\frac{\pi}{2}]$, the initial state $\ket{\phi}$ in Eq.~(\ref{eq:initialState}) interpolates between a separable and maximally entangled state. For this initial state we have
\begin{align}
	&r_3(t_{\rm int})=s_3(t_{\rm int})=\cos(2\eta),\cr
	&\chi_{11}(t_{\rm int})=-\chi_{22}(t_{\rm int})=\sin(2\eta),\cr
	&\chi_{33}(t_{\rm int})=1,
\end{align}
and rest of $r_i(t_{\rm int})$, $s_i(t_{\rm int})$  and $\chi_{ij}(t_{\rm int})$ are zero. Hence according to Eq.~(\ref{eq:rt}) for arbitrary time $t>t_{\rm int}$ we have
\begin{equation}\label{eq:rt2}
	\vec{r}(t)=\begin{pmatrix}
		0\\
		0\\
		\cos(2\eta)
	\end{pmatrix}
\end{equation}
For the rest of coefficients, we work in two different regimes in the following to subsections. For the regime of $v\to 0$ and $vt\ll\Delta$ we use analytical results in \S\ref{sec:TPME} and for other range of parameters we solve the master equation using Qutip library \cite{Qutip1, Qutip2}.
\subsection{Slow regime }
In this subsection we work in the regime of $v\to 0$ and $vt\ll\Delta$. We derive the explicit expression for entanglement when the initial state is a maximally entangled states. We discuss entanglement survival time and for zero temperature bath we discuss the behaviour of entanglement in terms of spin coupling direction to the environment. 

For initial state in Eq.~(\ref{eq:initialState}) by using Eq.~(\ref{eq:st}) for the slow-driving regime
we have 
\begin{equation}
   \vec{s}(t)=\begin{pmatrix}
    0\\
    0\\
    (\cos(2\eta)-\frac{a_-}{a_+})e^{-2a_+(t-t_{\rm int})}+\frac{a_-}{a_+}\\
    \end{pmatrix}.
   \label{eq:r3solu}
\end{equation}
Also, regarding Eq.~(\ref{eq:chit}) for initial state in Eq.~(\ref{eq:initialState}) the non-vanishing elements of matrix $\chi(t)$ have the following simple form:
\begin{align}
    &\chi_{11}(t)=-\chi_{22}(t)=e^{-b(t-t_{\rm int})}\cos{2k(t-t_{\rm int})}\sin{2\eta},\cr
    &\chi_{12}(t)=\chi_{21}(t)=-e^{-b(t-t_{\rm int})}\sin{2k(t-t_{\rm int})}\sin{2\eta},\cr
    &\chi_{33}(t)=e^{-2a_+(t-t_{\rm int})}\sin{2\eta}.
    \label{eq:chisolu}
\end{align}
From $\vec{r}(t)$, $\vec{s}(t)$ in Eq.~(\ref{eq:rt2}) and Eq.~(\ref{eq:r3solu}) and $\chi(t)$ in Eq.~(\ref{eq:chisolu}), we derive the density matrix $\tilde{\rho}_{SR}(t)$ which has X form. Hence its
partial transpose  $\tilde{\rho}_{SR}^{\top_R}(t)$ has X form and is given by
\begin{align}
\label{eq:rhoabt}
     \tilde{\rho}_{SR}^{\top_R}(t)
     &=\frac{1}{4}(\operatorname{id}\otimes\operatorname{id}+{s}_3(t)\sigma_z\otimes \operatorname{id}+r_3(t)\operatorname{id}\otimes \sigma_z\cr &+\sum_{i,j=1}^3(-1)^{j+1}\chi_{ij}(t)\sigma_i \otimes\sigma_j).
\end{align}
This block-diagonal structure enables us to derive the eigenvalues of $\tilde{\rho}_{SR}^{\top_B}(t)$ analytically:
\begin{align}
\label{eq:mu}
 \mu_{1,2}(t)&=\frac{1}{4}(1-\chi_{33}(t)\cr
 &\pm\sqrt{4(\chi^2_{11}(t)+\chi_{12}^2(t))+(s_3(t)-r_3(t))^2}),\cr
 \mu_{3,4}(t)&=\frac{1}{4}\left(1+\chi_{33}(t)\pm \abs{s_3(t)+r_3(t)}\right),
\end{align}
with $r_3(t)$, $s_3(t)$ and $\chi_{ij}(t)$ given in Eqs.~(\ref{eq:rt2}), (\ref{eq:r3solu}) and (\ref{eq:chisolu}). 
In order to see when the dynamics become entanglement breaking \cite{EB}, it is sufficient to consider a maximally entangled state as the initial state and investigate at what time the entanglement between two qubits vanishes \cite{EB}. Hence, in Eq.~(\ref{eq:initialState}) and subsequent equations we set $\eta=\frac{\pi}{4}$ for initial maximally entangled state. By examining eigenvalues of $\tilde{\rho}_{SR}^{\top_B}(t)$ in Eq.~(\ref{eq:mu}) we see that for $\eta=\frac{\pi}{4}$,  the only eigenvalue that might get negative values is $\mu_2(t)$. Hence, for a maximally entangled initial state and in the limit of $v\to 0$ and $t\ll\frac{\Delta}{v}$ negativity is given by
\begin{equation}
    \label{eq:NegInLimit}
    \mathcal{N}(\rho_{SR}(t))=\abs{\mu_2(t)}.
\end{equation}
When one share of an entangled pair, experience noise, the maximum entanglement survival time $\tau_{\rm ent}$ is defined as the smallest evolution time after which the dynamics becomes entanglement breaking \cite{Gatto2019}. Here in  the regime of $v\to 0$ and $t\ll\frac{\Delta}{v}$, we can find $\tau_{\rm ent}$ by equating Eq.~(\ref{eq:NegInLimit}) to zero for $\theta=0$:
\begin{equation}
\label{eq:ENT}
    \tau_{\rm{ent}}=-\frac{\ln{(\xi)}}{2b},
\end{equation}
where $b$ is defined in Eq.~(\ref{eq:apmb}) and
\begin{equation}
\label{eq:xi}
    \xi=\frac{3-\ell^2-2\sqrt{2-\ell^2}}{1-\ell^2}; \;\;\ell=\frac{1}{2\bar{n}(2\Delta)+1}
\end{equation}
\begin{figure}
    \centering
    \includegraphics[width=\columnwidth]{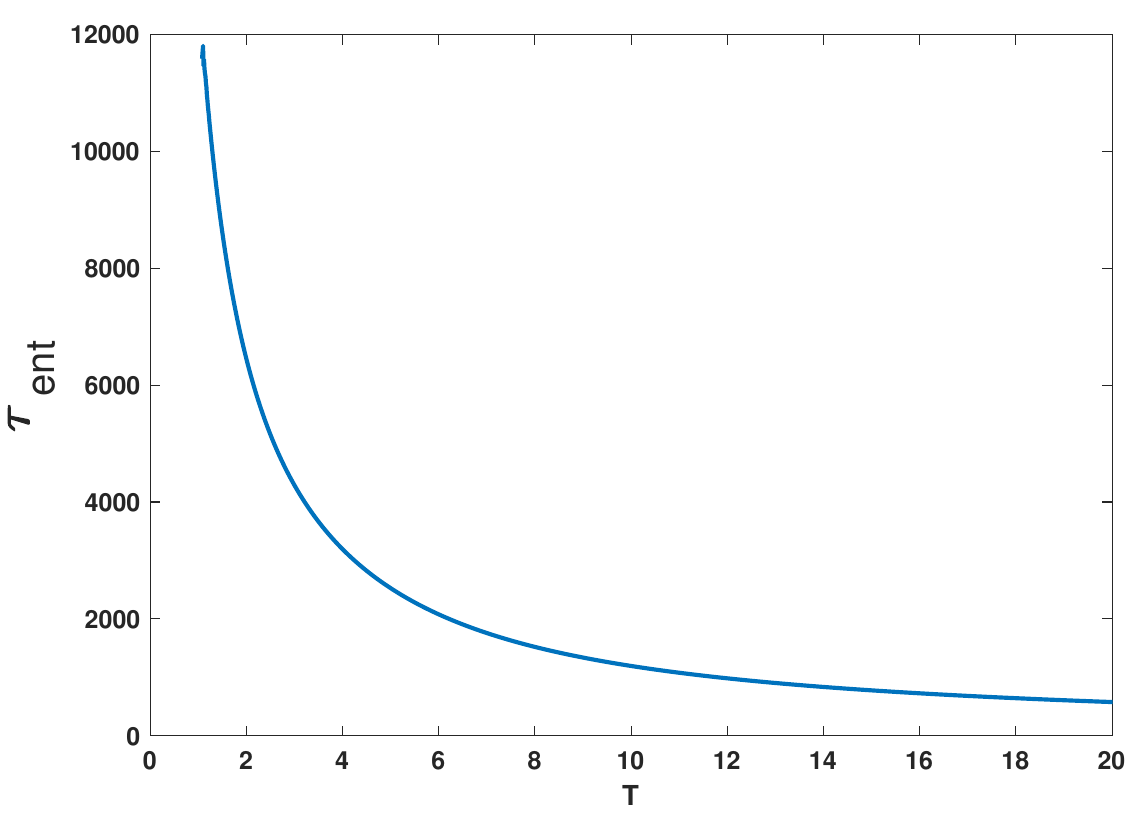}
    \caption{Entanglement survival time (Eq.~(\ref{eq:ENT})) versus temperature $T$ in slow deriving regime for $\theta=0$, $\lambda=0.1$, $\Delta=10$ and $\omega_c=\frac{\Delta}{3}$. }
    \label{fig:TauENT}
\end{figure}
Figure~(\ref{fig:TauENT}) shows $\tau_{\rm ent}$ in Eq.~(\ref{eq:ENT}) for $\theta=0$ versus temperature $T$ in the slow-driving regime when $\Delta=10$, $\omega_c=\Delta/3$, $\lambda=0.1$. This choice of parameters assures that the inequalities in Eq.~(\ref{eq:Secularapp}) are satisfied. As expected, temperature has destructive effect on $\tau_{\rm ent}$. What is reflected in Eq.~(\ref{eq:ENT}) and Fig.~(\ref{fig:TauENT}) is for $\theta=0$. For further investigation on the role of noise coupling direction on entanglement behaviour, we focus on zero temperature environment. By using Eq.~(\ref{eq:NegInLimit}) in the regime of  $v\to 0$ and $t\ll\frac{\Delta}{v}$ at $T=0$ we have
\begin{equation}
	\label{eq:NT0}
	\mathcal{N}(\rho_{SR}(t))=\frac{1}{2}e^{- (t-t_{\rm int})\lambda^2\pi J(2\Delta){\cos^2(\theta)} }. 
\end{equation}
It exhibits the exponential decay of entanglement in time and also provides an explicit relation between entanglement and  spin-coupling direction to the bath at $T=0$. 
\begin{figure}[h]
    \centering
    \includegraphics[width=\columnwidth]{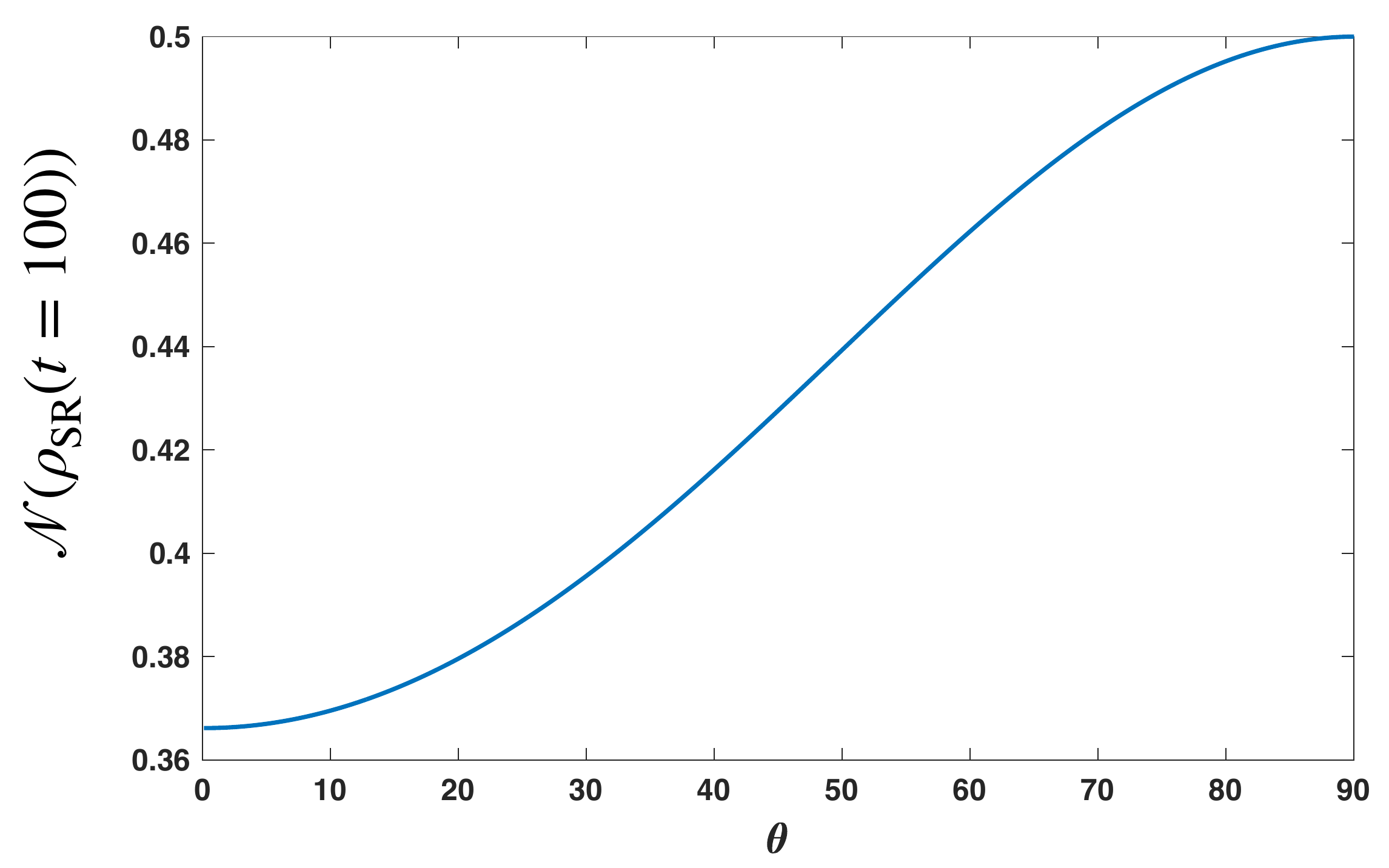}
    \caption{Entanglement versuse parameter $\theta$ (in degree) for $T=0$, $\Delta=10$, $t_0=-100$, $t=100$ as given in Eq.~(\ref{eq:NT0}).}
    \label{fig:NvsTheta}
\end{figure}
In  Fig.~(\ref{fig:NvsTheta}) we see entanglement between two qubits versus 
$\theta$ at $T=0$, for $t_{\rm int}=-100$, $t=100$ 
 $\Delta=10 $, 
 $\omega_c=\Delta /3$ and $\lambda=0.1$ as given in Eq.~(\ref{eq:NT0}). As entanglement  is non-increasing in time under local noise, from Fig.~(\ref{fig:NvsTheta}) we conclude that, at any instant of the evolution, for larger values of $\theta$, the noisy environment performs less destructively and at $\theta=\frac{\pi}{2}$ the initial entanglement is preserved. This is expected as in the regime of  $vt\ll\Delta$ and $v\to 0$, the system Hamiltonian in the rotated basis $\tilde{H}_S(t)$, is proportional to $\sigma_z$ (See  Eq.~(\ref{eq:tildeHLZ}) ). On the other hand by using Eq.~(\ref{eq:Lt}) and Eq.~(\ref{eq:kfgl}) in this regime we have:
\begin{equation}
\label{eq:Semi-AD}
\tilde{\mathcal{L}}_t[\bullet]=-i k[\sigma_z,\bullet]+\frac{\lambda^2}{4}\cos\theta^2(\sigma_-\bullet\sigma_+-\frac{1}{2}\{\sigma_+\sigma_-,\bullet\}),
\end{equation}
This is a representation of an amplitude damping channel when coupling to the bath is proportional to $\abs{\lambda \cos(\theta)}$. The larger value of $\theta$ is, the role of decoherence is weaker and the coherent terms plays the dominant role in the generator of the dynamics. At $\theta=\frac{\pi}{2} $, there is no dissipation term in the generator and coherent part leads to a unitary evolution which does not change entanglement. 
\subsection{fast-deriving regime}
In this subsection we analyse the behaviour of entanglement in the fast-deriving regime. We address its decay in time for different values of $\Delta$ and also discuss how it behaves in terms of the ratio $\frac{\Delta^2}{v}$, which characterises adiabatic and non-adiabatic regimes in the absence of interaction with any environment. 

In order to investigate the role of parameters affecting entanglement in fast-deriving regime, one should solve the differential equations with time-dependant coefficients in Eq.~(\ref{eq:hard1}). 

\begin{figure}
	\centering
	\includegraphics[width=1\linewidth]{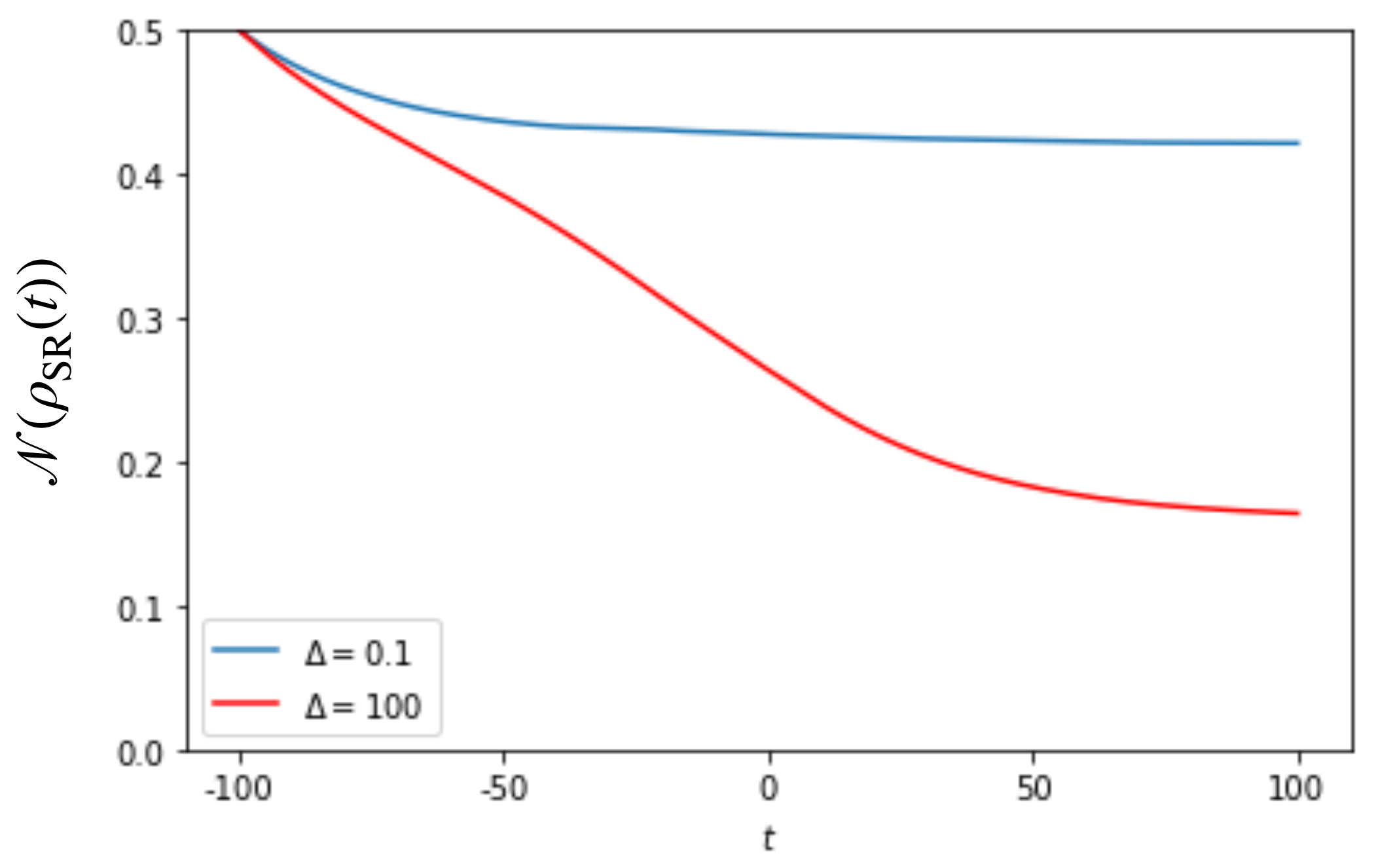}
	\caption{Negativity versus time for the initial state $\ket{\phi}=\frac{1}{\sqrt{2}}(\ket{00}+\ket{11})$, $\lambda=0.1$, $T=0$, $\omega_c=\frac{\Delta}{3}$, $\theta=0$ and $v=1$. From top to bottom  $\Delta=0.1, 100$. }
	\label{fig:diffv0}
\end{figure}

Unlike the limit of $v\to 0$ and $vt\ll\Delta$, analyticical results can not be obtained in the fast-deriving regime. Hence, we use Qutip library \cite{qutip} to solve the master equation and analyse entanglement behaviour numerically. The outcome of our code is in agreement with our analytical result in the range of $v\to 0$ and $vt\ll\Delta$. In Fig.(\ref{fig:diffv0}), the behaviour of negativity versus time is shown for initial state $\ket{\phi}=\frac{1}{\sqrt{2}}(\ket{00}+\ket{11})$,  $v=1$ and $\theta=0$, $\omega_c=\frac{\Delta}{3}$  and $T=0$. From top to bottom $\Delta=0.1,100$. As it is seen in this figure, for smaller value of $\Delta$ entanglement degradation is slower and after a shorter time it achieves its steady value. In other words, when $\frac{\Delta^2}{v}\ll 1$, that indicates a non-adiabatic regime, entanglement remains less intact by the environment compared to the adiabatic regime. Same behaviour is seen for $\theta=\frac{\pi}{2}$. 

\begin{figure}[h]
	\centering
	\includegraphics[width=\linewidth]{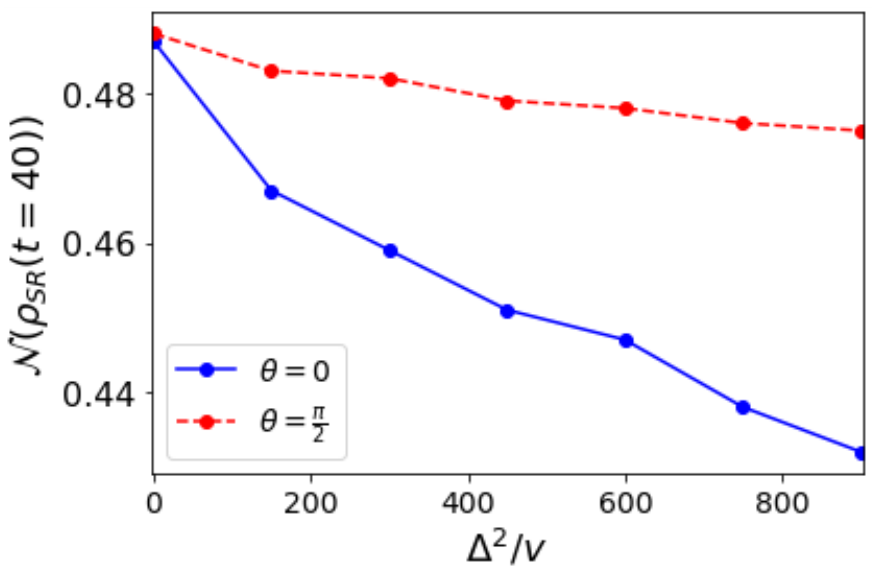}
	\caption{Negativity versus $\frac{\Delta^2}{v}$ for $\theta=0$ (blue curve) and $\theta=\frac{\pi}{2}$ (dashed red curve) at $t=40$ when initial state is a maximally entangled state, $t_0=-40$, $T=0$, $\omega_c=\frac{\Delta}{3}$, $\lambda=0.1$ and $v=1$.}
	\label{fig:NvsD2/v}
\end{figure}
To better illustrate
 the effect of adiabaticity on the dynamic on entanglement degradation, in Fig.~(\ref{fig:NvsD2/v}) we present the behaviour of entanglement versus the ratio $\frac{\Delta^2}{v}$ for $\theta=0$ (blue solid curve) and $\theta=\frac{\pi}{2}$ (red dashed curve) at $t=40$. Like previous cases, the initial state is a maximally entangled state. As it is seen in this figure, the larger the value of $\frac{\Delta^2}{v}$ the entanglement decay is stronger. Therefore, non-adiabatic regime is more favorable for preserving entanglement.

\section{Conclusion}\label{sec:conclusion}

We have studied entanglement degradation when one share of an entangled pair undergoes a dissipative Landau-Zener evolution. Regarding the role  of Landau-Zener model in adiabatic quantum computation \cite{Farhi2000, Aharonov2007} and its success in describing quantum annealing \cite{Santoro2006, Arceci2017}, the dissipative Landau-Zener model serves as a realistic model for modelling local noise and analysing entanglement degradation. 
This study has many applications in different  areas such as communication and quantum network developments \cite{kimble_quantum_2008, wehner_quantum_2018, Pompili2021}. At the same time this is a  challenging problem because the Landau-Zener Hamiltonian is time-dependent. 
Hence, considering the Markovian dissipative dynamic for modelling the noise, requires a careful derivation of master equation and choice of parameters to guarantee the validity of approximations made in derivation of Markov master equation. Furthermore, the ultimate generator of the dynamics, is time-dependent which makes it complicated to have analytical solution for the master equation in its general form is not possible. 

After careful derivation of master equation using the approach in \cite{ogawa}, to overcome the complications for solving coupled differential equations  with time-dependent coefficients, first we concentrate in the slow-deriving regime with $v\to 0$ and $vt\ll\Delta$. In this regime
when noise is longitudinal  we drive an analytical expression for entanglement survival time which confirms the destructive role of bath temperature on entanglement. Also we present analytical expression for entanglement when bath temperature is zero. Our analytical results in the slow-deriving regime, shows the important role of spin-coupling direction to the environment in entanglement preservation. In \cite{Arceci2017} 
role of spin-coupling direction on the ground-state probability is discussed and the  beneficial effect of bath temperature on ground-state probability is seen only for transversal noise ($\theta=\frac{\pi}{2}$). Here,
we have another aspect of the importance of spin-coupling direction. In the slow-deriving regime and in presence of bath with zero temperature, by increasing $\theta$  the destructive effect of environment on entanglement decreases and for transversal coupling, entanglement is totally preserved. Role of spin-coupling direction is also evident in the fast-driving regime where we solve the master equation numerically. As it seen in Fig.~(\ref{fig:NvsD2/v}) for all values of $\frac{\Delta^2}{v}$ entanglement between two qubits is larger for transversal noise compared to longitudinal noise. 

Another important question is about the role of adiabaticity of the dynamic on entanglement degradation. Despite the primary naive expectation, our results shows that in the non-adiabatic regime, entanglement degradation is less compared to the adiabatic evolution. In other words, for fixed value of minimum gap $\Delta$, for larger values of $v$ the entanglement degradation is smaller.

While the case of two level system has its own importance and impact, a promising direction to follow would be to extend this study to systems with higher dimensions. In systems with higher dimension, in addition to adiabatic property of the dynamics and, role of spin coupling-directions, one can  study different classes of time-dependent Landau-Zener type Hamiltonians \cite{Kiselev2013}. This would give us more knowledge about the factors affecting entanglement degradation.

\section*{Acknowledgement}
Authors acknowledges financial support by Sharif University of Technology, Office of Vice President for research under Grant No. G930209.  LM
acknowledges 
support from the ICTP through the Associates Programme (2019-2024).

\appendix
\section{Kossakowski coefficients}
\label{sec:coef}

In this appendix we explain how to compute the coefficients $\gamma(\omega_m(t))$ in Eq.~(\ref{eq:Kossakowski}) and $\mathcal{S}(\omega_m(t))$ in Eq.~(\ref{eq:LS}).
These coefficients are given in terms of $\Gamma(\omega_m(t))$:
\begin{equation}
\label{eq:Gamma}
\Gamma(\omega_m(t))=\int_{0}^\infty d\tau\int_{-a}^ad\nu e^{i(\omega_m(t)-\nu)\tau} \operatorname{Tr}\left(\textbf{B}(\nu)\textbf{B}\rho_E^{\rm th}\right)
\end{equation}
where $a$ is determined such that all transition frequencies belong to the interval $(-a,a)$ and
\begin{equation}
\label{eq:Bw}
    \textbf{B}(\nu):=\int_{0}^{\omega_{\rm{max}}}d\mu \ket{\mu}\bra{\mu}\textbf{B}\ket{\mu+\nu}\bra{\mu+\nu},
\end{equation}
with $\ket{\mu}$ representing the eigenstate of $b^\dagger(\mu)b(\mu)$ with eigenvalue $\mu$. The coefficient $\gamma(\omega_m(t))$ is the real part of $\Gamma(\omega_m(t))$. With straightforward calculations \cite{ogawa} we have 
\begin{equation}
\label{eq:Gengamma}
\gamma(\omega_m(t))=2\pi\operatorname{Tr}\left(\textbf{B}(\omega_m(t))\textbf{B}\rho^{\rm{th}}_E\right),
\end{equation}
By using the definition in Eq.~(\ref{eq:Bw}) and definition of thermal state in Eq.~(\ref{eq:Bw}), the more explicit form of $\gamma(\omega_m(t))$ in Eq.~(\ref{eq:Kossakowski}) is derived.  The coefficients $\mathcal{S}(\omega_m(t))$ in Eq.~(\ref{eq:LS}) are given by the imaginary part of $\Gamma(\omega_m(t))$ in Eq.~(\ref{eq:Gamma}) \cite{ogawa}.
\bibliography{EntanglementDegradationDLZ}

\end{document}